\documentclass{elsart}

\usepackage{amsfonts,amsmath,amssymb,bm,cite}

\usepackage{amssymb}

\begin{document}

\begin{frontmatter}


 \title{Current-conserving nonlinear response theory in driven systems}

 \author{J. Peralta-Ramos}
 \address{CONICET and Departamento de F\'isica, Facultad de Ciencias Exactas y Naturales, Universidad de Buenos Aires-Ciudad Universitaria, Pabell\'on I, 1428 Buenos Aires, Argentina}
 \ead{jperalta@df.uba.ar}

 \author{E. Calzetta}
 \address{CONICET and Departamento de F\'isica, Facultad de Ciencias Exactas y Naturales, Universidad de Buenos Aires-Ciudad Universitaria, Pabell\'on I, 1428 Buenos Aires, Argentina}
 \ead{calzetta@df.uba.ar}

\begin{abstract}
Employing the closed-time path 2PI effective action (CTP 2PI EA) approach, we study the response of an open interacting electronic system to time-dependent external electromagnetic fields. We show that the 2PI EA provides a systematic way of calculating the propagator and response functions of the system. Due to the invariance of the 2PI EA under external gauge transformations, the response functions calculated from it are such that the Ward-Takahashi hierarchy, that ensures current conservation beyond the expectation value level, is satisfied. These findings may be useful in the study of interacting electronic pumping devices, and serve to clarify the connection between current conservation (beyond the mean value level) and real-time nonlinear response theory. 
\end{abstract}

\end{frontmatter}

\section{Introduction}
\label{intro}
 
In this work we determine the basic requirements that a field-theoretical approach to open and driven systems must satisfy in order to produce current conserving results (in the sense of the Ward-Takahashi hierarchy developed in Sec. \ref{wth}), in transport calculations {\it going beyond linear response}. Another aim we have in mind is to clarify the close relation that exists between current conservation and response theory, especially in the nonlinear regime. In order to analyze these issues, we combine the so-called external gauge invariance method with the closed-time path two-particle irreducible  coarse-grained effective action (CTP 2PI CGEA), which has proved a valuable tool in the study of strongly interacting quantum open systems both in and out of equilibrium \cite{libro,calgauge}. Although our analysis is quite general and can be applied to a variety of condensed matter systems, we have in mind a system of {\it strongly} interacting electrons subjected to external driving fields and in contact with two ideal reservoirs of noninteracting electrons. This could be a picture of the so-called interacting electron pumping devices, which are currently attracting much interest both experimentally and theoretically \cite{pump}.

\section{Ward-Takahashi hierarchy}
\label{wth}
In many-particle systems, $n$-point vertex functions represent generalized currents which satisfy the hierarchy of Ward-Takahashi identities \cite{rivier}. This hierarchy is satisfied to all orders {\it in the exact theory}, warranting local gauge invariance and the conservation of the associated charges. 
In its most general form, the generalized continuity equation (WTH) can be written as (we use the simplified notation $1=(t_1,\bf{r_1})$ and employ Schwinger-Keldysh non-equilibrium formalism \cite{kel,chou,libro}): 
\begin{equation}
\begin{split}
& \partial_\mu^z \Lambda^\mu_{(n)}(1\ldots n,1'\ldots n';z) = \\
& i^n e \{ -\delta(z-n')G_n(1\ldots n,1'\ldots (n-1')z)+\ldots \\
& +(-1)^n \delta(z-1')G_n(1\ldots n,2'\ldots z)+\ldots \\
& +\delta(z-n)G_n(1\ldots (n-1)z,1'\ldots n')-\ldots \\
& -(-1)^n\delta(z-1)G_n(2\ldots z,1'\ldots n')\}
\end{split}
\label{jer}
\end{equation}
where 
\begin{equation}
\begin{split}
& \Lambda^\mu_{(n)}(12\ldots n,1'2'\ldots n';z) = \\
& <T_c j^\mu(z)\psi(1)\psi(2)\ldots\psi(n)\psi^\dagger(n')\ldots \psi^\dagger(2')\psi^\dagger(1')>
\end{split}
\label{lam}
\end{equation}
is the $(n+1)$-point vertex function with current insertion at $z=(t_z,\bf{r_z})$,  

\begin{equation}
\begin{split}
& j^\mu(z)=-e \lim_{z' \rightarrow z} D^\mu(z,z')\psi^\dagger(z')\psi(z) \qquad ;\\
& D^i(z,z') = (2i)^{-1}(\nabla^i_z-\nabla^i_{z'} \\
&\qquad \qquad -ie[A^i(z)+A^i(z')])\qquad \mu=i=1,2,3 \\
& D^0(z,z') = 1 \qquad \qquad \qquad \qquad \qquad \qquad \mu=0
\end{split}
\label{curr}
\end{equation}
with the external fields denoted by $A_\mu$, and $G_n$ are real-time propagators defined as usual 
\begin{equation}
\begin{split}
& G_{n}(1\ldots n,1'\ldots n') = \\
& i^{-(n)}<T_c \psi(1)\ldots \psi(n)\psi^\dagger(n')\ldots\psi^\dagger(1')> \qquad .
\end{split}
\label{gn}
\end{equation}
The classical continuity equation and the usual WT identity correspond to the cases $n$=0 and $n$=1 of Eq. (\ref{jer}), respectively.

As the WTH shows, in the exact theory particle number is {\it strongly} conserved, not only in the mean. This means that for a many-particle system in the presence of an external field, the current is conserved to all orders in the external perturbation. 

It is impossible, in general, to obtain the propagators of an interacting field theory exactly, and some approximations must be made. In particular, for a strong coupled theory re-summation concepts are usually needed \cite{libro,dedom}, and warranting conservation laws then becomes a nontrivial issue.  
A systematic way of generating conserving approximations (at the classical level, i.e. $n$=0) was given by Baym \cite{baym} and corresponds to his well-known $\Phi$-derivable scheme. The solutions obtained from truncating the $\Phi$ functional are such that the expectation values of the respective Noether currents are conserved \cite{baym,libro}. We emphasize that this situation corresponds to the case $n$=0 of the WTH. Therefore, the conservation of generalized currents encoded in the WTH is not automatically warranted in this approach. 

Although in some cases the conservation of current at the expectation value level may be sufficient, it is not suitable for nonlinear response studies. 
The external gauge method \cite{bando} is a possible way of overcoming this problem present in $\Phi$-derivable approaches\footnote{Another method is the re-summation technique due to Hees and Knoll \cite{hees}, but we shall not discuss it here.}. It provides a systematic way of generating consistent Scwhinger-Dyson (SD) and Bethe-Salpeter (BS) equations that automatically satisfy WT identities. Most importantly in the context of transport theories of strongly interacting systems, the derivation of the SDE and the BSEs can be done to any order in the external field coupled to the system. Therefore, the latter is especially suited for the study of response theory beyond first order and its relation to current conservation. 

\section{CTP 2PI effective action}
\label{ctp}
 
The system we are interested in consists of a central region with interacting electrons described by fields $(\psi^\dagger,\psi)$ and coupled to an external field. This central region is connected to two ideal reservoirs of noninteracting electrons described by fields $(\phi_\alpha^\dagger,\phi_\alpha)$, with $\alpha=(L,R)$ denoting left or right reservoir. The reservoirs are assumed to remain in equilibrium at all times. 
The CTP classical action \cite{libro} of the system is $S[\bar\psi,\psi,\bar\phi,\phi] = S_\psi + S_\phi + S_c$, with:
\begin{equation}
\begin{split}
S_\psi &= c_{AB}\bar\psi^A C^{-1}_{AB} \psi^B + S_{int}[\bar\psi,\psi] \\
S_\phi &= c_{AB}\bar\phi^A B^{-1}_{AB} \phi^B \\
S_c &= c_{AB} \bar\psi^A T_{AB} \phi^B + \textrm{h. c.}
\end{split}
\label{claction}
\end{equation}
where 
\begin{equation}
S_{int}[\bar\psi,\psi] = \frac{1}{24} U_{ABCD} \bar\psi^A \psi^B \bar\psi^C \psi^D 
\label{sint}
\end{equation}
being $U_{ABCD}$ the completely antisymmetrized bare interaction local vertex. $T_{AB}$ is a local coupling parameter between the central region and the reservoirs. 

We are using a DeWitt notation \cite{libro} with $A=(x,a)$, $x=(t_x,r_x,\sigma)$ and $a=(1,2)$ [or $(+,-)$] being CTP indices indicating the branch within the closed-time contour. For the fields describing electrons inside the reservoirs, an additional index $\alpha$ must be included in the CTP indices $(A,B)$, but for simplicity we leave it implicit. Repeated indices are assumed to be integrated or summed. $c_{ab}$ is a CTP metric $c_{ab}=\textrm{diag}(1,-1)$, while $c_{AB} = c_{ab}\delta(x_A,x_B)$. $C_{AB}$ and $B_{AB}$ are the free CTP propagators corresponding to $S_\psi$ and $S_\phi$. They satisfy equations of motion governed by single-particle Hamiltonians
\begin{equation}
\begin{split}
\tilde{h}_0(1) &= \frac{1}{2}[-i\nabla_1 + e A_i(1)]^2 + eA_0(1) \\
h_0(1)&= -\frac{1}{2}\nabla^2_1  \qquad ,
\end{split}
\end{equation} 
respectively.
By performing the double Legendre transform on the generating functional of connected propagators, and then using the background field method we can write the CTP 2PI CGEA of the system as \cite{libro}
\begin{equation}
\Gamma[G,A] = i \textrm{Tr ln} G - i D^{-1}_{AB} G^{AB} +\Gamma_2[G] \qquad ,
\label{gamma}
\end{equation}
where $D_{AB}^{-1}=C_{AB}^{-1}+i\Sigma_{\phi,AB}$, being $\Sigma_{\phi,AB}$ the self-energy due to the reservoirs. 
$\Gamma_2[G]$ encodes all quantum corrections and consists of vacuum 2PI closed diagrams with full propagators in internal lines and vertices corresponding to a theory with shifted classical action $S[\hat{\bar\psi}+\bar\psi,\hat\psi+\psi]$ (neglecting constant and linear terms), where $(\bar\psi,\psi)$ denote fluctuations. The diagrams are vacuum because the mean value of the fluctuation fields is zero by construction. 
Note that because of the vanishing of the mean fields, the shifted classical action vanishes when evaluated at $(\hat{\bar\psi},\hat\psi)$. The case of noninteracting electrons corresponds in this scheme to $\Gamma_2=0$.
  
The Schwinger-Dyson equations for the CTP propagators follow directly from $\Gamma[G,A]$
\begin{equation}
\frac{\delta \Gamma}{\delta G_{AB}} = -\frac{1}{2} K_{AB}
\end{equation}
where the physical case corresponds to vanishing external sources $K = 0$. Defining the self-energy 
\begin{equation}
\Sigma_{AB}[G]=-2 \delta \Gamma_2[G,A]/\delta G^{AB}
\end{equation}
we can rewrite the SDE in the usual way
\begin{equation}
G_{AB}^{-1} = D^{-1}_{AB} +i\Sigma_{AB}[G] \qquad ,
\end{equation}
being $\Sigma_{AB}$ one-particle irreducible by construction. 

We note that in the 2PI EA approach, mean fields and two-point propagators are treated on the same footing, whereas higher order propagators are obtained from them \cite{libro}. Although this represents an approximation to the exact dynamics of the quantum system, we will show that it is consistent with the WTH. 
 
\section{WT hierarchy from external gauge invariance}
\label{wtegisec}

Under a local transformation $U(1)=\exp{ie\varphi(1)}$, the external field and the full propagator transform as \cite{rei,calgauge} (we omit the CTP indices for the moment)
\begin{equation}
\begin{split}
& A_\mu \rightarrow A_\mu' = U A_\mu U^{-1} - i(\partial_{\mu} U) U^{-1}  \\
& G(1,2) \rightarrow G'(1,2) = U(1)G(1,2)U^{-1}(2) \qquad .
\end{split}
\label{egi}
\end{equation}
The crucial observation that allows us to relate the CTP 2PI CGEA of the open system to nonlinear transport through it is that $\Gamma[G,A]$ is invariant under a gauge transformation of the external field $A_\mu$. Following Bando, Harada and Kugo (Ref. \cite{bando}), we will call this {\it external gauge invariance} (EGI) of the 2PI EA. It is rather straightforward to prove that, if we retain all terms in the loop expansion of $\Gamma_2$, then the 2PI CGEA is EGI \cite{libro,rei,calgauge}. This also holds if we truncate the loop expansion of $\Gamma_2$ at arbitrary order \cite{j}. 

We will now make a connection between EGI and nonlinear response. For notational simplicity, CTP indices are omitted in the following. The solution $G[A]$ to the SDE can be expanded in powers of the external field $A_\mu$ (see Refs. \cite{chou,heinz}): 
\begin{equation}
\begin{split}
&G[A](X,Y) = G[0](X,Y) + iA_\mu \Pi_3^\mu + \\
&\frac{i^2}{2} A_\mu A_\nu \Pi_4^{\mu\nu} + \frac{i^3}{3}A_\mu A_\nu A_\rho \Pi_5^{\mu \nu \rho} \ldots \\
&= \sum_{n=0} \frac{i^n}{n!} \int d1\ldots dn A_{\mu_1}(1)\ldots 
A_{\mu_n}(n) \\
& \Pi_{(n+2)}^{\mu_1\ldots \mu_n}(1,\ldots,n; X,Y) \qquad ,
\end{split}
\label{expans}
\end{equation}
where in the second line we have made explicit the internal (integration) $(1,\ldots,n)$ and the external $(X,Y)$ variables. 
The ``response'' functions $\Pi_{(n+2)}$ encode the variation of the full propagator with the external field. 
It is worth remarking that the use of the closed-time path method automatically ensures that the response functions are causal \cite{libro,chou}. 

Returning now to Eq. (\ref{expans}), the response functions are given by
\begin{equation}
\begin{split}
& \Pi_{(n+2)}^{\mu_1\ldots \mu_n}(X,Y;1\ldots n) = \\
&(-i)^n<T_c j^{\mu_1}(1)\ldots j^{\mu_n}(n) \psi(X)\psi^\dagger(Y) > 
\end{split}
\label{pidef}
\end{equation}
and correspond to $(n+2)-$point functions with $n$ current vertices inserted at locations $(1,\ldots,n)$ where interactions between the current and the external classical field take place. 
The functions $\Pi_{(n+2)}$ are obtained from the SDE by functional differentiation with respect to $A_\mu$ (and then setting $A=0$). This results in the BSEs for the response functions.

The combination of the SDE and the BSEs completely determine the full propagator and the response functions $\Pi_{(n+2)}$. The SDE is obtained from the 2PI CGEA, while the BSEs are obtained from the SDE by differentiation with respect to the external field. The important point is that, because the 2PI CGEA is invariant under external gauge transformations, both the full propagator and response functions obtained this way are external gauge covariant. As we will show below, the EGI property of the 2PI CGEA implies that $G$ and $\Pi_{(n+2)}$, as obtained from $\Gamma[G,A]$, satisfy the WT hierarchy. This provides the required link between current conservation in nonlinear response and the external gauge invariance of the 2PI CGEA, and also a powerful and systematic way of studying nonlinear response in strongly interacting systems coupled to ideal reservoirs.

To see the connection between EGI and the WT hierarchy, recall that external gauge invariance of the effective action means $G[A']=UG[A]U^{-1}$. Inserting the expansion in powers of the external field given in Eq. (\ref{expans}) into both sides of this identity we get
\begin{equation}
\begin{split}
& G[0] + i A'_\mu \Pi_3^\mu + \frac{i^2}{2}A'_\mu A'_\nu \Pi_4^{\mu \nu} + \ldots =\\
& UG[0]U^{-1} + iA_\mu U\Pi_3^\mu U^{-1} + \frac{i^2}{2}A_\mu A_\nu U\Pi_4^{\mu \nu} U^{-1} + \ldots 
\end{split}
\label{egiward}
\end{equation}

In particular, for an infinitesimal external gauge transformation $U(X)\approx 1 + ie\varphi(X)$ the transformed external field is 
\begin{equation}
A'_\mu(X) = A_\mu(X) + \partial_\mu \varphi(X) \qquad ,
\end{equation}
so Eq. (\ref{egiward}) becomes
\begin{equation}
\begin{split}
& \sum_{n=0} \frac{i^n}{n!} \int [dn]\prod_{i=1}^n \{A_{\mu_i}(i)+\partial_{\mu_i}\varphi(i)\}\Pi_{(n+2)}^{\mu_1\ldots \mu_n} =\\
& \sum_{n=0} \frac{i^n}{n!} \int [dn]\prod_{i=1}^n A_{\mu_i}(i)\{\Pi_{(n+2)}^{\mu_1\ldots \mu_n}+ie[\varphi(X)\Pi_{(n+2)}^{\mu_1\ldots \mu_n}-\\
& \Pi_{(n+2)}^{\mu_1\ldots \mu_n}\varphi(Y)]\}
\end{split}
\label{produc}
\end{equation}
where we have suppressed the arguments of $\Pi_{(n+2)}^{\mu_1\ldots \mu_n}$ and defined $[dn]=d1\ldots dn$ for brevity. Comparing terms of the same order in $A_\mu$ on both sides of this expression we get
\begin{equation}
\begin{split}
& \Pi_2(X,Y)+ i \int d1 [\partial_\mu^1 \varphi(1)]\Pi_3^\mu(X,Y;1) =\\
& \Pi_2(X,Y) + ie [\varphi(X)\Pi_2(X,Y)-\Pi_2(X,Y)\varphi(Y)]
\end{split}
\label{zerotha}
\end{equation}
for the zeroth order term. Higher order terms have a similar structure but can become quite involved \cite{j}. The main point to emphasize is that the EGI of the 2PI CGEA implies relationships among the response functions, shown explicitly in Eq. (\ref{produc}).

Assuming that $\varphi$ vanishes at infinity, we can integrate by parts the second term in Eq. (\ref{zerotha}) (zeroth order in $A_\mu$) to obtain 
\begin{equation}
\begin{split}
& i\int d1 \varphi(1)\partial_\mu^1 \Pi_3^{\mu}(X,Y;1) =\\
& -i e[\varphi(X)\Pi_2(X,Y)-\Pi_2(X,Y)\varphi(Y)]
\end{split}
\end{equation}
which implies
\begin{equation}
\begin{split}
& \partial_\mu^z <T_c j^\mu(z)\psi(X)\bar\psi(Y)> = \\
& e<T_c \psi(X)\bar\psi(Y)> [\delta(Y-z)-\delta(X-z)] \qquad .
\end{split}
\label{zeroward}
\end{equation}
This is precisely the identity corresponding to $n=1$ in the WT hierarchy given by Eq. (\ref{jer}). We see that, even at zeroth order in the external field, the relation between the three-point vertex and the two-point function, Eq. (\ref{zeroward}), is satisfied due to the EGI of the 2PI EA. 

It is clear that this procedure could be continued to higher order terms in $A_\mu$, thus generating higher order WT identities. We note that the hierarchy obtained for the response functions $\Pi_{(n+2)}$ is completely equivalent to that involving $\Lambda_{(n)}$, given in Eq. (\ref{jer}), as expected since, ultimately, they both enforce current conservation \cite{j}. 

\section{Conclusions}
\label{concsec}

We have determined the basic requirements that an approximation to a non-equilibrium many-body problem in an open and driven fermionic system must satisfy in order to achieve current conservation beyond the expectation value level. One of the most important results of this work is the close relation found between nonlinear response theory and the Ward-Takahashi hierarchy, necessary for current conservation. This connection was clearly displayed by using an approximation scheme based on the CTP 2PI CGEA, that provides equations for the propagator and response functions consistent with the WTH. We emphasize that, although being an approximation to the true dynamics of the system, the 2PI EA description is consistent with the WTH, even in the nonlinear regime. 
Our results may be of use in the theoretical study of quantum transport through interacting electronic pumping devices, which are nowadays receiving much attention.  

We thank Liliana Arrachea, Carlos Na\'on, Alfredo Levy Yeyati and Sergio Ulloa for useful discussions. This work has been supported in part by ANPCyT, CONICET and UBA (Argentina).

\end{document}